\documentclass{article}
\usepackage{color}

\usepackage{xargs}
\usepackage{graphicx}
\usepackage[acronym]{glossaries}
\usepackage{subfig}
\usepackage{wrapfig}
\usepackage{placeins}
\usepackage{tikz}
\usepackage{array}
\usepackage{listings}
\usepackage{pifont}
\usepackage{colortbl}
\usepackage{arxiv}

\usepackage[utf8]{inputenc} 
\usepackage[T1]{fontenc}    
\usepackage{hyperref}       
\usepackage{url}            
\usepackage{booktabs}       
\usepackage{amsfonts}       
\usepackage{nicefrac}       
\usepackage{microtype}      
\usepackage{lipsum}
\usepackage{graphicx}

\usepackage{lipsum,multicol}

\newacronym{AM}{AM}{App Modeler}
\newacronym{CM}{CM}{Configuration Model}
\newacronym{ABM}{ABM}{App Behavior Model}
\newacronym{ABMM}{ABMM}{App Behavior Meta-Model}
\newacronym{GM}{GM}{Generator Modeler}
\newacronym{GMo}{GMo}{Generation Model}
\newacronym{AG}{AG}{App Generator}
\newacronym{NA}{NA}{Native App}

\newacronym{POI}{POI}{Point of Interest}

\newacronym{QR}{QR}{Quick Response}
\newacronym{CMS}{CMS}{Content Management System}
\newacronym{IDE}{IDE}{Integrated Development Environment}

\newacronym{MDE}{MDE}{Model-Driven Engineering}
\newacronym{MDA}{MDA}{Model-Driven Architecture}
\newacronym{BPM}{BPM}{Business Process Management}
\newacronym{BPMS}{BPMS}{Business Process Management System}
\newacronym{BPMN}{BPMN}{Business Process Model and Notation}
\newacronym{GPS}{GPS}{Global Positioning System}

\newacronym{MAML}{MAML}{M\"unster App Modeling Language}
\newacronym{DSL}{DSL}{Domain-Specific Language}

\newacronym{REST}{REST}{Representational State Transfer}
\newacronym{API}{API}{Application Programming Interface}
\newacronym{B2E}{B2E}{Business to Employee}

\newacronym{Sass}{Sass}{Syntactically Awesome Stylesheets}
\newacronym{CSS}{CSS}{Cascading Stylesheets}

\newacronym{RWD}{RWD}{Responsive Web Design}
\newacronym{HTML}{HTML}{Hypertext Markup Language}
\newacronym{JS}{JS}{JavaScript}

\newacronym{WAR}{WAR}{Web Application Archive}
\newacronym{JPA}{JPA}{Java Persistence API}
\newacronym{MVC}{MVC}{Model–View–Controller}

\newacronym{EMF}{EMF}{Eclipse Modeling Framework}
\newacronym{DOM}{DOM}{Document Object Model}

\newacronym{AmI}{AmI}{Ambient Intelligence}

\newacronym{ABR}{ABR}{Abstract Binding Repository}

\newacronym{FI}{FI}{Flow Instance}
\newacronym{BPEL}{BPEL}{Business Process Execution Language}
\newacronym{NDSPE}{NDSPE}{Native Domain-Specific Process Engine}

\newacronym{DSPML}{DSPML}{Domain-Specific Process Modelling Language}

\newcommand{\ac}[1]{\gls{#1}}

\newcommand{\acp}[1]{\Glspl{#1}}
\makeglossaries

\title{Modeling Support for Domain-Specific Application Definition}

\author{
	José Miguel Pérez-Álvarez \\
	Naver Labs Europe\\
	6 Chemin de Maupertuis\\
	Meylan \\
	\texttt{jm.perez@naverlabs.com} \\
	\And
	Adrian Mos \\
	Naver Labs Europe\\
	6 Chemin de Maupertuis\\
	Meylan \\
	\texttt{adrian.mos@naverlabs.com} \\

}

\begin{document}
	\maketitle
	
	\begin{abstract}
		
		In this paper we present the modeling support infrastructure for domain-specific application definition. This consists of a set of meta-models and the associated generators to allow the definition of reusable and domain-specific behavior blocks, which can later be used to compose complex behaviors. 
		
		In addition we also present the related visual languages that facilitate the creation of these models.
		

	\end{abstract}

\section{Introduction}
\label{sec:introduction}

\par The advantages of domain-specific approaches \cite{kelly2008domain} as well as the use of \acp{DSL} \cite{van2000domain} to effectively deal with application domains in software development \cite{mernik2005and} have been largely discussed \cite{articleDSLSurvey}. In this paper we present an extension to the models presented in \cite{mos2016generating, mos2016business, DBLP:conf/er/CornaxPML17}, that allows the modeling of high-level functions, through smaller domain-specific building blocks. 

\par In the literature we can find many domain-specific languages, which allow to model behavior in various contexts. With regard to business process modeling, the Business Process Model and Notation \cite{omg2011bpmn} (BPMN 2.0) has become the de-facto standard. With the aim at filling the Business-IT gap, significant effort has been put into bringing BPMN executable and closer to Service Oriented Architectures (SOA). While these components help alleviate agility problems that business stakeholders encounter, we observed that most of the existing solutions are domain-independent and platform-dependent, which limit the power of business matter experts at the design and monitoring stages.

\par The original use of the models we extend in this paper was related to moving from domain-specific descriptions to execution through transformations to \ac{BPMN} \cite{mos2016generating, mos2016business}, with the aim of leveraging the execution stacks of many available \ac{BPMS}. However such approaches have the following drawbacks: they requires technical involvement from business analysts as they must deploy the generated \ac{BPMN} models into the \ac{BPMS} and configure a number of parameters; The semantics of domain-specific processes don't map directly over \ac{BPMN} constructs and this entails the generation of additional  activities and third party components \cite{DBLP:conf/er/CornaxPML17}, increasing the complexity of the resulting artifacts; they introduce a dependency to an additional language and more importantly to a full execution stack that needs to be acquired and managed separately; they don't have mechanisms to isolate client-side code from changes on the server-side, impacting both sides during maintenance, with the disadvantages described above.

\par
\begin{figure}[h]
	\centering
	\includegraphics[width=0.5\textwidth]{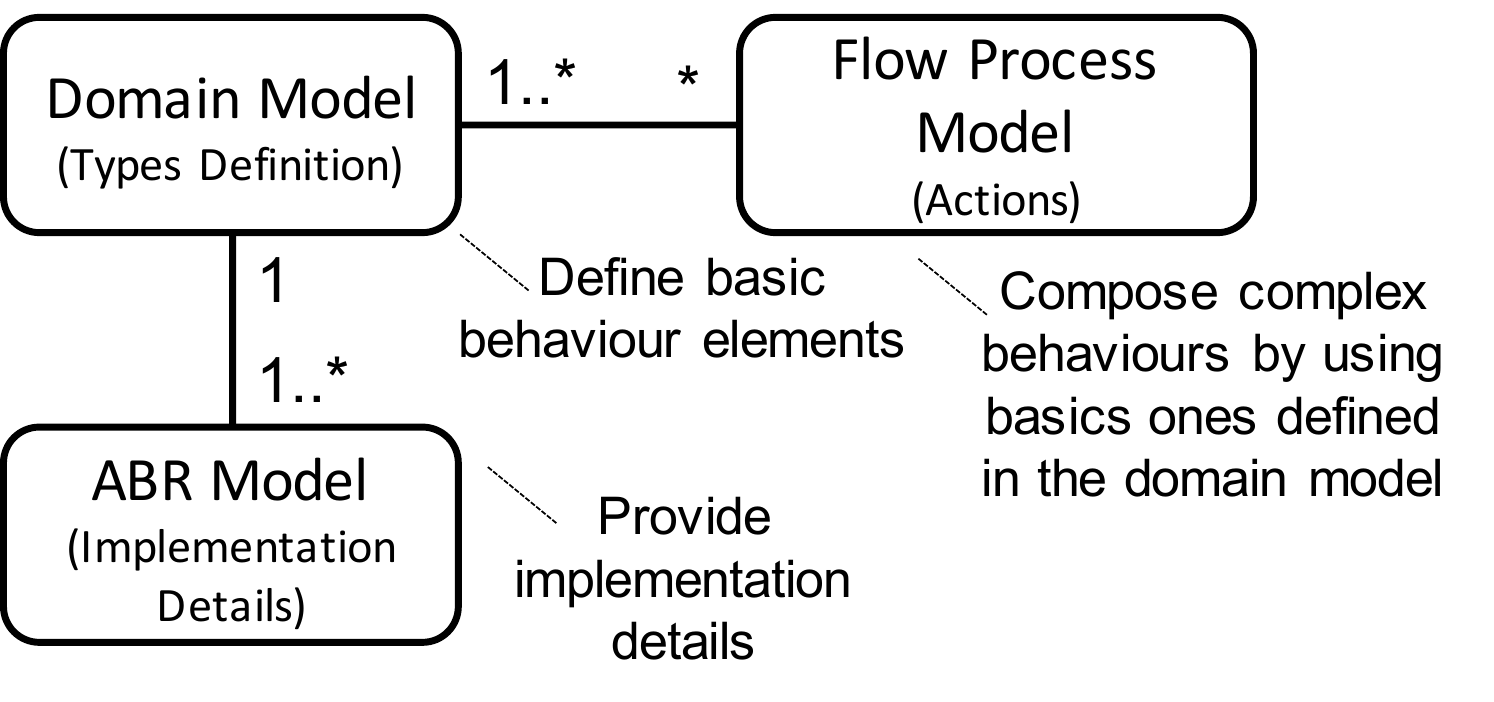}
	\caption{Relation of models}
	\label{fig:elements_involves}
\end{figure}

 The extension proposed on this paper introduces the necessary modeling concepts so that they can be interpreted by an engine without the need for external elements. Figure \ref{fig:elements_involves} shows graphically the different modeling blocks involved, as well as the relationship between them. The domain, \ac{ABR} meta-model and flow meta-models are adapted and reused from the approach presented in \cite{mos2016generating}. They have been modified and updated from their original form and these changes are highlighted in the text. Their instances, the domain, \ac{ABR}, and flow models respectively, are used when defining the behavior.

\par With the aim of illustrate the different modeling block, as well as the expressibility of them, in this paper we show how \textit{Conduit} can modeled with the proposed elements. \textit{Conduit} is a clone of medium.com, described in the RealWorld project \cite{Sun2019}. This large open-source project is effectively a programming benchmark providing many different implementations of the same app, using different technology stacks.

\par A \textbf{domain model} can be considered as a \ac{DSPML}, or in a more restricted sense, a domain-specific library. Its basic behavioral blocks, called \textit{Activities}, and the corresponding domain meta-model are presented in Section \ref{subsec:domain_meta_model}. The \textbf{\ac{ABR} model} contains technical configuration elements that relate to connections between the domain and a variety of services. The \ac{ABR} meta-model and an example are detailed in Section \ref{subsec:abr_meta_model}. A \textbf{flow model} can be seen as a reusable function that specifies complex behavior by composing different activities defined in one or several domain models. The details of this meta-model are detailed in Section \ref{subsec:process_meta_model}.

\section{The Domain Meta-Model}
\label{subsec:domain_meta_model}

\par
\begin{figure*}[t]
	\centering
	\includegraphics[width=0.65\textwidth]{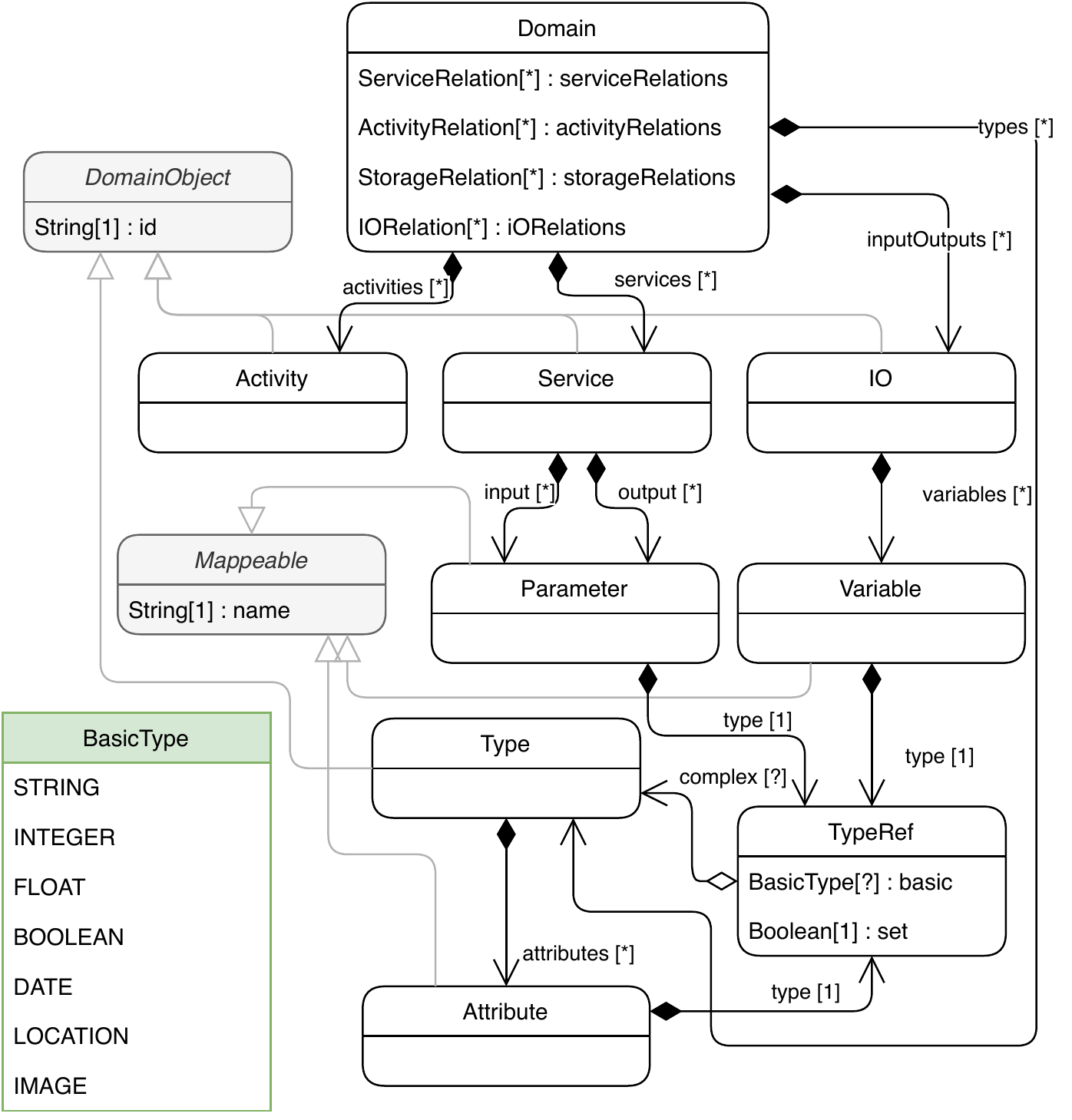}
	\caption{Domain Meta-Model - Element View}
	\label{fig:domain_metamodel_core}
\end{figure*}

 The elements that can be modeled in a domain can be seen in Figure \ref{fig:domain_metamodel_core}. The top-level element is \textit{Domain}. and it contains a set of \textit{Activity}, \textit{Service}, \textit{IO} and \textit{Type} elements.

\par A data \textit{Type} represents a kind of data structure. It is composed of a set of \textit{Attributes} that define the type. An \textit{Attribute} is defined by its id (it is a \textit{DomainObject}), and its type of data, identified by a \textit{TypeRef} element. A \textit{TypeRef} is a reference which only points to a type, without defining it. A \textit{TypeRef} can represent either a \textit{BasicType} or a \textit{complex} one, or indeed a set of such elements if its attribute \textit{set} is true. The modeling of Types as well as the referencing mechanism have been introduced in this paper and added to the original meta-model.

\par As can be seen in Figure \ref{fig:domain_metamodel_core}, the \textit{BasicTypes} that can be used in a domain are: STRING, INTEGER, FLOAT, BOOLEAN, DATE, LOCATION and IMAGE.

\par A \textit{Service} represents an entity that can be executed either by calling a function available in the system, or an external call.  Note that this is an abstract service, which must be connected to the concrete implementation through the \ac{ABR} model \cite{Mos:2013:PMD:2570455.2570912}.

\par An \textit{IO} element represents an input/output operation, i.e. information that must be retrieved from an external source or information that must be returned (for instance to be presented to a human). It is composed by a set of \textit{Variables}. The \textit{IO} does not specify whether the information to be retrieved must come from a human, or whether data to be exposed must be presented in a graphical user interface. For instance if the true location is requested, a mobile application could get it automatically from the device sensors, while a browser-based application could ask the end user to indicate it on a map.

\par
\begin{figure}[t]
	\centering
	\includegraphics[width=0.55\textwidth]{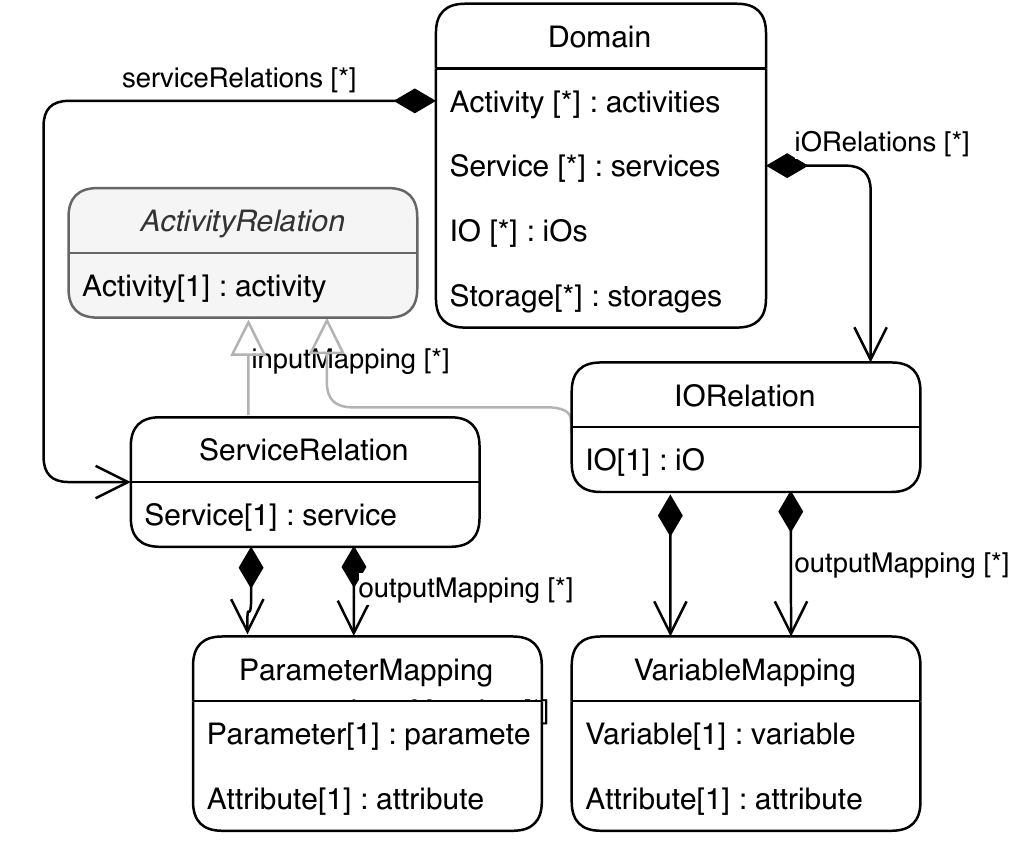}
	\caption{Domain Meta-Model - Relationship View}
	\label{fig:domain_metamodel_relations}
\end{figure}

 Finally, an \textit{Activity} represents an atomic unit of basic behaviour. As can be seen in Figure \ref{fig:domain_metamodel_relations}, the domain meta-model contains relations between \textit{Activities} and the elements defined above, through a series of \textit{ServiceRelations} and \textit{IORelations}.

\par A \textit{ServiceRelation} relates an \textit{Activity} (it is a subtype of \textit{ActivityRelation}) to a \textit{Service}, and also defines a mapping between the input/outputs parameters of the service and the attributes of a \textit{Type}. At runtime there is a conceptual space managed by the execution engine, where activities exchange information, called the \textbf{data-flow}. The activities read, create or update values from the data-flow.

\par In the same way, an \textit{IORelation} relates an \textit{Activity} to an \textit{IO} element, and it also models the source of the information to publish from the data-flow (\textit{outputMappings}), as well as how the data-flow is updated with the requested information (\textit{inputMappings}).

\par The following changes have been provided in this version of Domain meta-model compared to its original incarnation:

\begin{itemize}
	\item Concept of \textit{IO}: This replaces a similar element called Form, that was used to model rendering policies. The new approach gives complete freedom in the representation offered to the clients of the execution element.
	
	\item Complex type support: The original version supports only basic types, while the new one adds multiplicity and brings increased reuse by allowing the definition of attributes to point to other domain types.
	
	 \item New basic types: LOCATION and IMAGE were not part of the original version.
	
	\item Data-flow mappings: The mappings (\textit{ParameterMapping} and \textit{ValueMapping}) relate to data-flow access, which is a local data store. This allows more efficient data exchange between tasks in flows, compared to the alternative of persisting everything in an external database. It also allows for better monitoring support.
	
	\item Database support: In the original version, \textit{Entities} and \textit{Relationships} were specifically modeled \cite{DBLP:conf/er/CornaxPML17}, now they can be automatically derived from the \textit{Types}.
\end{itemize}

\par
\begin{figure*}[t]
	\centering
	\includegraphics[width=0.95\textwidth]{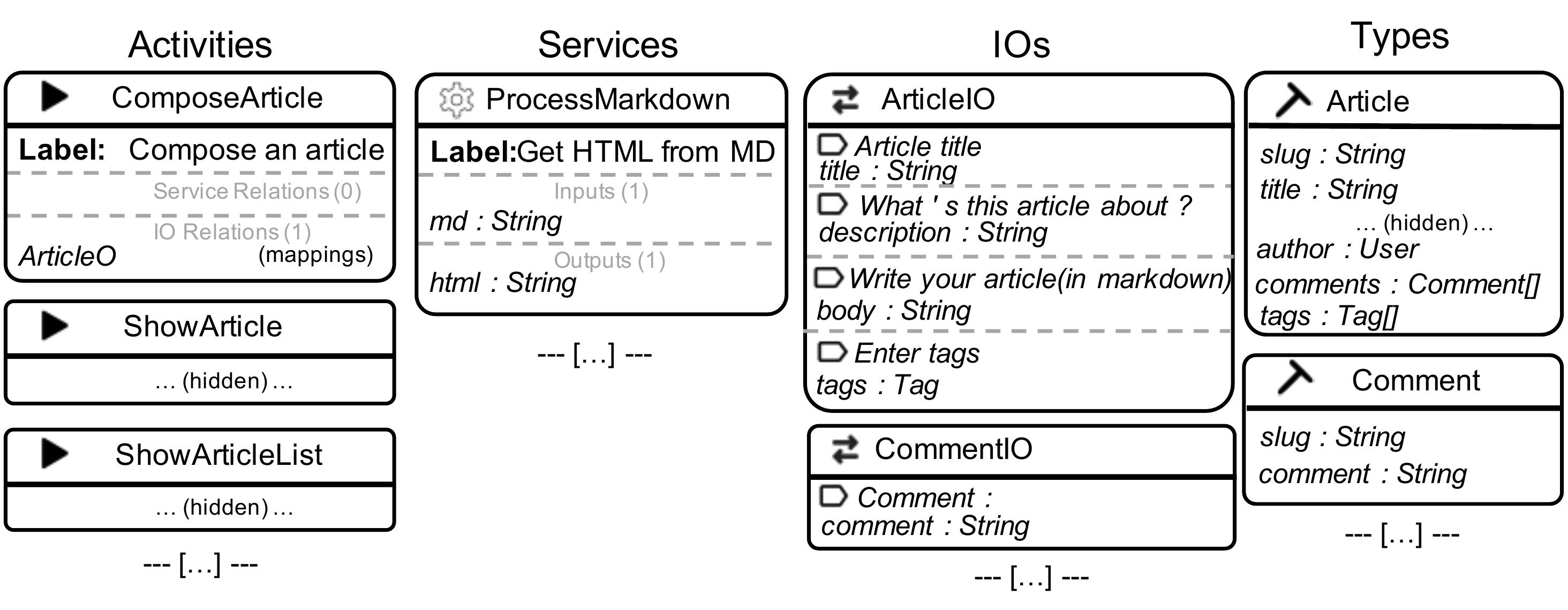}
	\caption{Domain Sample}
	\label{fig:domain_sample}
\end{figure*}

 Figure \ref{fig:domain_sample} shows a sample domain model using a graphical representation (the current implementation of the approach supports both textual and graphical modeling of domains). This corresponds to the \textit{Conduit} example mentioned in Section \ref{sec:introduction}, but only contains basic specific behavioral elements focused on Articles.

\par These elements include two types: \textit{\textbf{Article}} that contains the actual textual information; \textit{\textbf{Comment}} that represents an opinion about an article. Regarding \textit{Services}, only one is defined, called \textit{\textbf{ProcessMarkdown}}, used to transform a given text in Markdown format to HTML, in order to be rendered in a webpage. Some IOs can also seen in the Figure \ref{fig:domain_sample}: \textit{\textbf{ArticleIO}}, to obtain information about the article that a user want to post, and \textit{\textbf{CommentIO}} to obtain comments about an article. Finally there are three activities, \textit{\textbf{ComposeArticle}} to ask users to compose an article, \textit{\textbf{ShowArticle}} to provide article details to the end-user, lastly \textit{\textbf{ShowArticleList}} to retrieve and return a list of articles.

\section{ABR Meta-Model}
\label{subsec:abr_meta_model}

\par
\begin{figure}[t]
	\centering
	\includegraphics[width=0.5\textwidth]{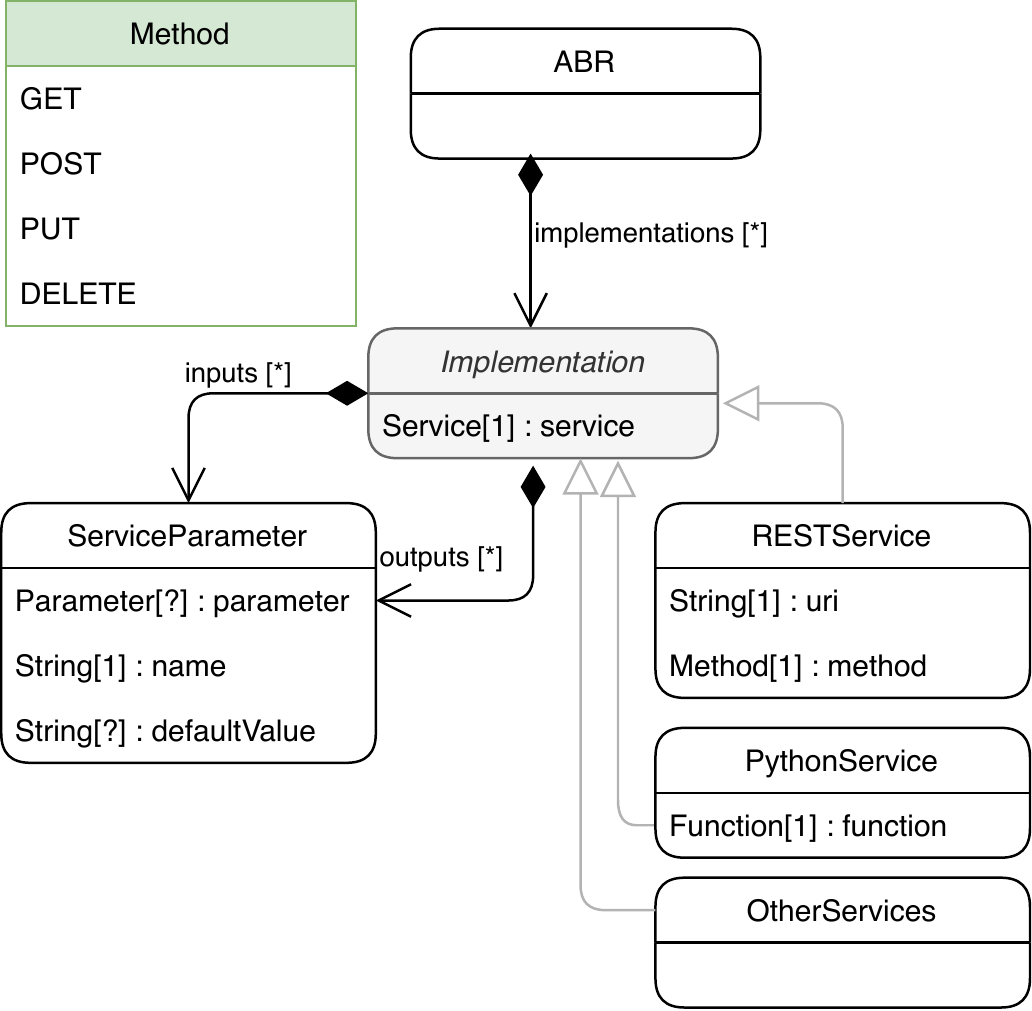}
	\caption{ABR Meta-Model}
	\label{fig:abr_metamodel_relations}
\end{figure}

 The \ac{ABR} is presented in Figure \ref{fig:abr_metamodel_relations}. The \textit{Services} modeled in the domain are abstract services.

\par The \textit{Implementation} meta-class is composed of implementation details as well as a set of \textit{ServiceParameters} as inputs and outputs. The \textit{ServiceParameter} element models the name of the real variables of the service implementation, and maps them with the \textit{Parameters}.

\par Finally, as illustrated in Figure \ref{fig:abr_metamodel_relations}, the \textit{Implementation} meta-class is abstract. The reason is that we can model various type of services, ranging from a \textit{PythonService} that represents a script written in Python, to a \textit{RESTService} that represents the integration with a third-party system by using \ac{REST}, to any (\textit{OtherServices}). The \ac{ABR} meta-model is therefore prepared to be extended with new \textit{Implementation} types.

\par
\begin{figure}[t]
 	\centering
 	\includegraphics[width=0.55\textwidth]{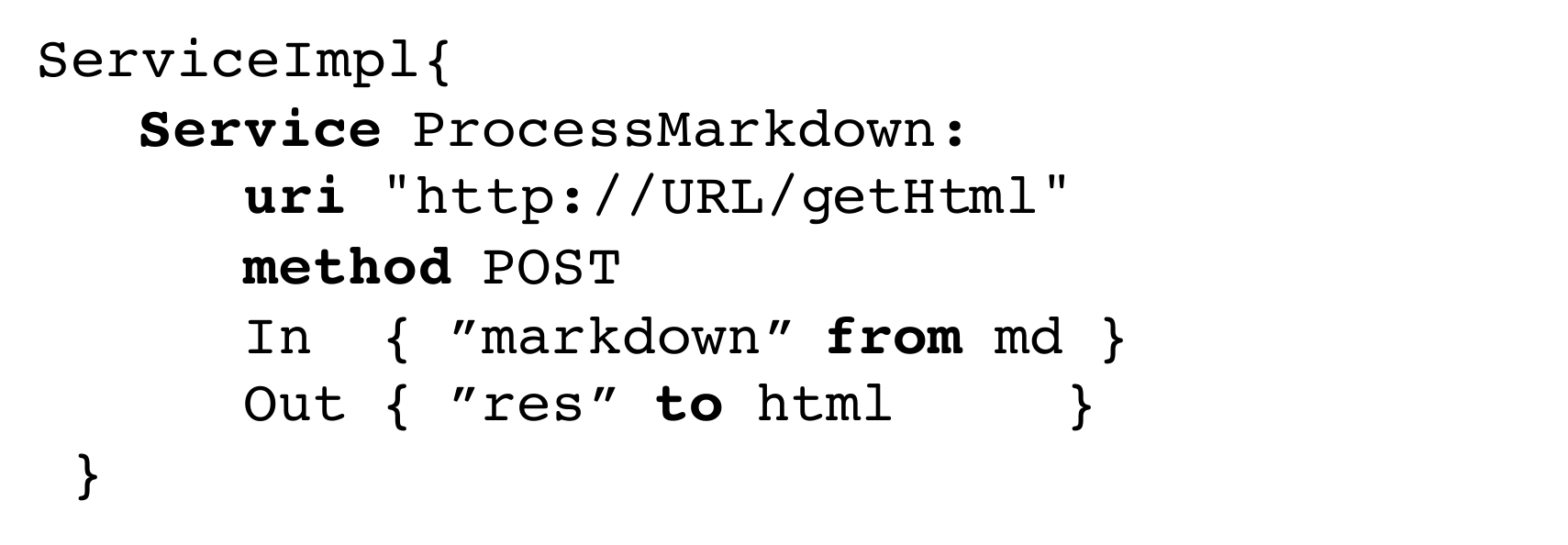}
 	\caption{ABR sample}
 	\label{fig:abr-sample}
 \end{figure}

 Figure \ref{fig:abr-sample} shows an example of an \ac{ABR} model, in textual form. As can be seen, it is providing implementation details for the service \textit{ProcessMarkdown}, which in this case is exposed through a POST \ac{REST} call. 
 
\section{Flow Meta-Model}
\label{subsec:process_meta_model}

\par
\begin{figure*}[t]
	\centering
	\includegraphics[width=0.65\textwidth]{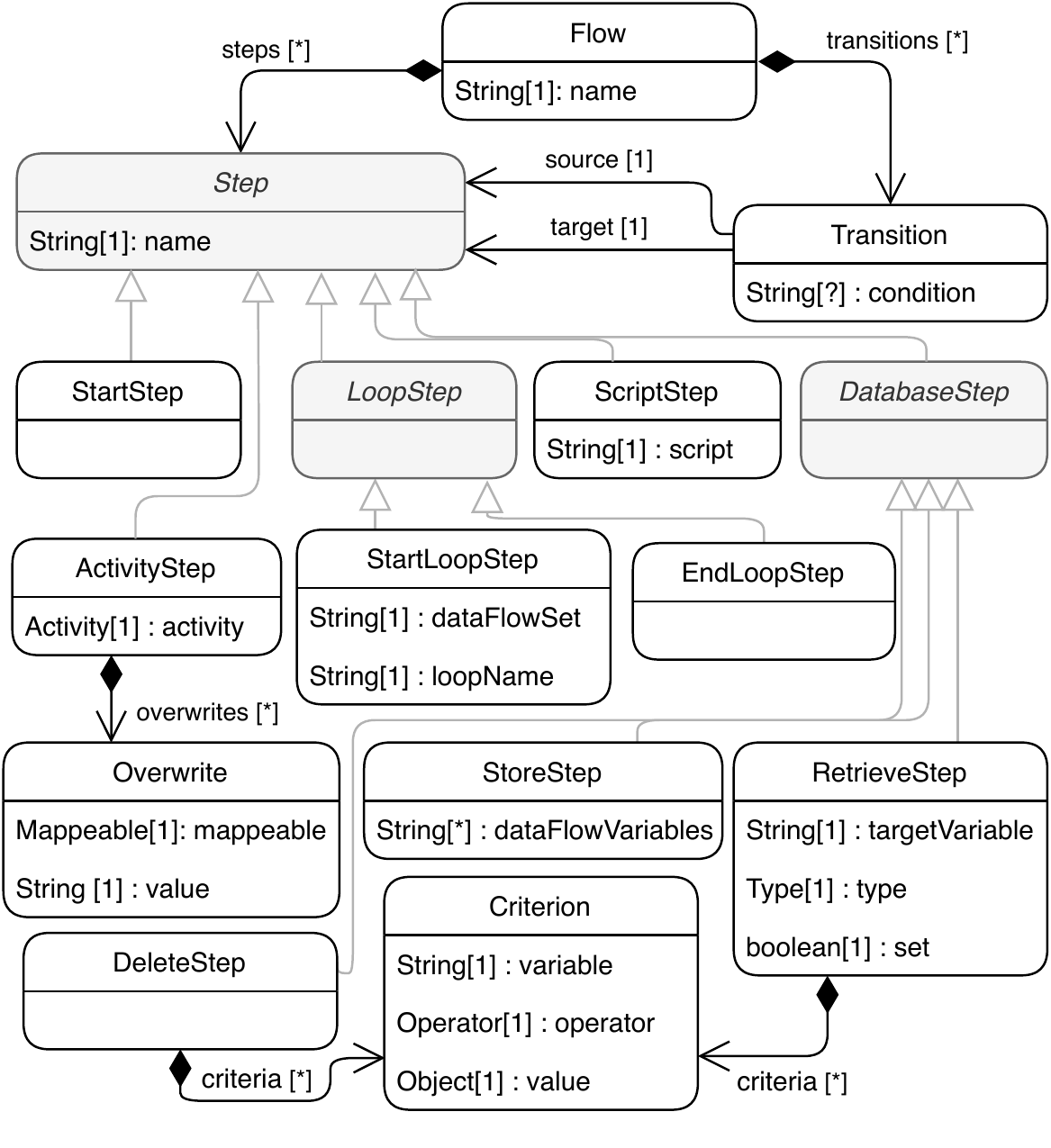}
	\caption{Flow Meta-Model}
	\label{fig:proces_metamodel_relations}
\end{figure*}

 A \textbf{\textit{Flow}} models a function of behavior by combining activities defined in a set of domains. As can be seen in Figure \ref{fig:proces_metamodel_relations}, it is composed of \textit{Steps} and associated \textit{Transitions}. On one hand, a \textit{Step} represents an action that must be performed. On the other hand, a \textit{Transition} (with or without \textit{conditions}) establishes a relationship between  \textit{source} and \textit{target} \textit{Steps}.

\par Furthermore, a \textit{Transition} establishes the flow of information between two \textit{Steps}. This \textbf{defines the data-flow}, which has the effect that the information generated in the source Step will be available to be used in the target Step, and any subsequent steps as reachable through the transition graph.

\par The \textit{Step} meta-class is abstract, therefore cannot be instantiated. There are five different types of \textit{Steps}: \textit{StartStep}, \textit{ActivityStep}, \textit{LoopStep}, \textit{ScriptStep}, and \textit{DatabaseStep}.

\par A \textit{StartStep} is optionally used to indicate to the engine the execution starting point in case of possible ambiguity (that can occur in certain situation such as when some loops don't leave out obvious first steps). Naturally, \textit{Steps} of type \textit{StartStep} cannot be target of any \textit{Transition}.

\par An \textit{ActivityStep} represents the usage in a flow of domain (\textit{Activities}). The inputs and outputs defined in the respective activities can be set at design time, effectively overwriting any values they might otherwise hold at runtime. For instance, a variable called "user-message" could be set to a certain value by the application designer that wishes at a certain stage in the execution that a given message be displayed. To this end an \textit{ActivityStep} contains a set of \textit{Overwrite} elements.

\par \textit{LoopSteps} are used to iterate over collections of variables existing in the data-flow at runtime. \textit{LoopStep} is abstract, and has two sub-types, \textit{StartLoopStep} and \textit{EndLoopStep} modeling the beginning and end of the loop respectively. \textit{StartLoopStep} is composed of the name of the set to iterate on (\textit{dataFlowSet}) as well as the name of the loop in data-flow (\textit{loopName}) at runtime.

\par A \textit{ScriptStep} allows creation of operations such as creation and assignment of variables or addition of elements to sets. This type of \textit{Step} can also be used to match certain functionally similar, but syntactically different variables, in cases where several compatible domains are used for the same flow.

\par A \textit{DatabaseStep} is used to connect with databases. It is abstract and the available sub-types are \textit{StoreStep}, \textit{RetrieveStep} and \textit{DeleteStep}. The fist one represents an operation to persist a set of variables (\textit{dataFlowVariables}) existing in the data-flow at runtime. The second one is used to bring data from a database into variables in the data-flow. This \textit{RetrieveStep} is composed of the name of the variable where the resulting data must be stored (\textit{targetVariable}), the data type to retrieve (\textit{type}), a boolean indicating if the resulting variable is a collection (\textit{set}), as well as a set of \textit{Criterion} used to specify the retrieval \textit{criteria}. The meta-class \textit{Criterion} represents a condition, and it is composed of a \textit{variable}, an \textit{operator} and a \textit{value}. The last sub-type, the \textit{DeleteStep}, represents an operation to remove data from the database, using the same selection semantics as the \textit{RetrieveStep}. Operations to update objects from database can be performed by combining retrieve and store steps.

\par
\begin{figure*}
	\centering
	\includegraphics[width=0.8\textwidth]{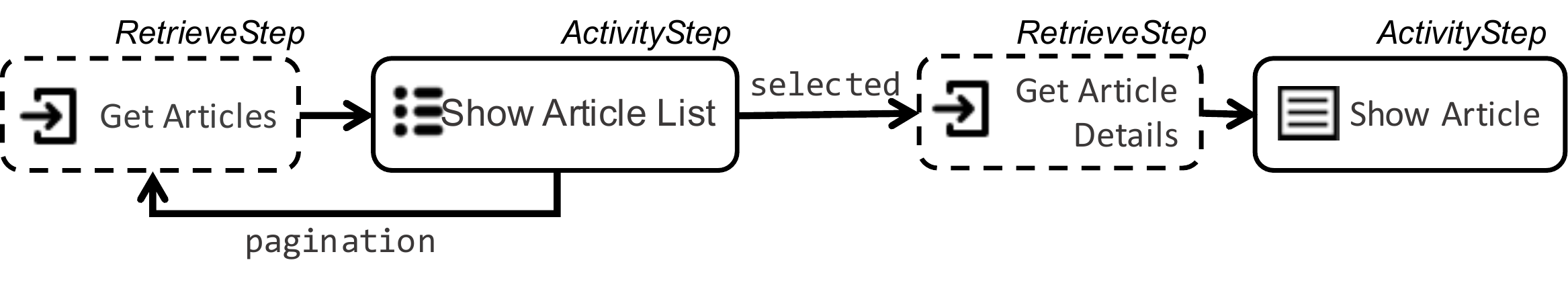}
	\caption{Flow Model}
	\label{fig:proces_model}
\end{figure*}

 A flow example can be seen Figure \ref{fig:proces_model} where each rectangle represents a \textit{Step} for the execution of a domain activity, while each arrow models a \textit{Transition}. The first step, \textit{Get articles}, is a \textit{RetrieveStep} and its purpose is to retrieve a the set of articles stored in a database. In the next step, the articles are shown (\textit{Show article list}) and the user select one of them, or chooses a new page listing articles (pagination), that corresponds to the activity \textit{ShowArticleList} in the domain presented in Figure \ref{fig:domain_sample}. Once an article is selected, its details are obtained in the \textit{RetriveStep} (\textit{Get article details}) and the full contents are shown to the user in the last step (\textit{Show article}).

\section{Conclusion and Future Work}
\label{sec:conclusions}

\par In this paper we present an approach that allow you to define small reusable blocks of behavior called domains, that can be used to model more complex behaviors called flows.

\par One of the possible advantages derived of this approach is that it is possible to define a division of roles, so that people with technical knowledge can define domains, which can be used with people with less technical knowledge, to model behavior in the form of flows. Therefore is possible to empower a variety of users to create specifications of application behavior.

\par As future work we want to evaluate the level of real knowledge that is necessary to model behavior using flows, and thus know if people without technical knowledge can create behavior by themselves. On the other hand, we also plan to create an execution engine that can interpret flow natively.

\bibliography{our-papers.bib}
\bibliographystyle{abbrv}

\end{document}